\documentclass[aps,prl,twocolumn,preprintnumbers,amsmath,amssymb,superscriptaddress]{revtex4}

\usepackage{graphicx}
\usepackage{pifont} 
\usepackage{wasysym} 
\usepackage{tabularx}
\usepackage{multirow}
\usepackage{hhline}

\newcommand{\Tr}{\mathrm{Tr}}

\begin{document}

\title{Rolling quantum dice with a superconducting qubit}

\author{R. Barends}
\thanks{These authors contributed equally to this work.}
\author{J. Kelly}
\thanks{These authors contributed equally to this work.}
\affiliation{Department of Physics, University of California, Santa
Barbara, CA 93106, USA}
\author{A. Veitia}
\affiliation{Department of Electrical Engineering, University of
California, Riverside, CA 92521, USA}
\author{A. Megrant}
\affiliation{Department of Physics, University of California, Santa
Barbara, CA 93106, USA}
\author{A. G. Fowler}
\affiliation{Department of Physics, University of California, Santa
Barbara, CA 93106, USA} \affiliation{Centre for Quantum Computation
and Communication Technology, School of Physics, The University of
Melbourne, Victoria 3010, Australia}

\author{B. Campbell}
\author{Y. Chen}
\author{Z. Chen}
\author{B. Chiaro}
\author{A. Dunsworth}
\author{I.-C. Hoi}
\author{E. Jeffrey}
\author{C. Neill}
\author{P. J. J. O'Malley}
\author{J. Mutus}
\author{C. Quintana}
\author{P. Roushan}
\author{D. Sank}
\author{J. Wenner}
\author{T. C. White}
\affiliation{Department of Physics, University of California, Santa
Barbara, CA 93106, USA}

\author{A. N. Korotkov}
\affiliation{Department of Electrical Engineering, University of
California, Riverside, CA 92521, USA}
\author{A. N. Cleland}
\author{John M. Martinis}
\affiliation{Department of Physics, University of California, Santa
Barbara, CA 93106, USA}

\date{\today}

\begin{abstract}
One of the key challenges in quantum information is coherently
manipulating the quantum state. However, it is an outstanding
question whether control can be realized with low error. Only gates
from the Clifford group -- containing $\pi$, $\pi/2$, and Hadamard
gates -- have been characterized with high accuracy. Here, we show
how the Platonic solids enable implementing and characterizing larger
gate sets. We find that all gates can be implemented with low error.
The results fundamentally imply arbitrary manipulation of the quantum
state can be realized with high precision, providing new practical
possibilities for designing efficient quantum algorithms.
\end{abstract}

\maketitle

The Platonic solids have been studied since ancient times for their
beauty and symmetry \cite{plato}, and make excellent random number
generators \cite{dnd3}. Here, we exploit their symmetry for quantum
information. Quantum processing would benefit from having a large set
of accurate gates to reduce gate count and error
\cite{hastings,lanyon2011,hanneke}, yet it is an open question
whether arbitrary gates can be implemented with low error -- only the
restricted group of Clifford gates \cite{kitaev1997,ross2014} has
been used with high precision \cite{barends2014,ryan2009,harty2014}.
We use the Platonic solids as a pathway and implement gate sets
inspired by the tetrahedron, octahedron, and icosahedron, including
gates never previously benchmarked. We achieve low error for all
gates. These results illustrate the potential of using unitaries with
a fine distribution, and suggest arbitrary rotations can be realized
with high accuracy, opening new avenues for performing gates and
designing algorithms efficiently.

Recently, major advances have been made in accurately implementing
octahedral (Clifford) gates on a variety of platforms.
Superconducting qubits, liquid NMR and ion traps have shown
single-qubit gate errors ranging from $10^{-3}$ to $10^{-6}$
\cite{barends2014,ryan2009,harty2014}, determined via Clifford-based
randomized benchmarking (RB). However, process verification of
non-Clifford gates is a conundrum: Quantum process tomography can be
used, but state preparation and measurement error can lead to
significant systematic deviations, limiting precision. Clifford-based
RB is insensitive to these errors, but unavailable for gates which
fall outside of the Clifford group. Here, the use of other rotational
groups allows for randomized benchmarking of non-Clifford gates. A
different approach to estimating errors of non-Cliffords was proposed
in Ref.~\cite{kimmel2014}.

\begin{figure}[b!]
    \centering
    \includegraphics[width=0.48\textwidth]{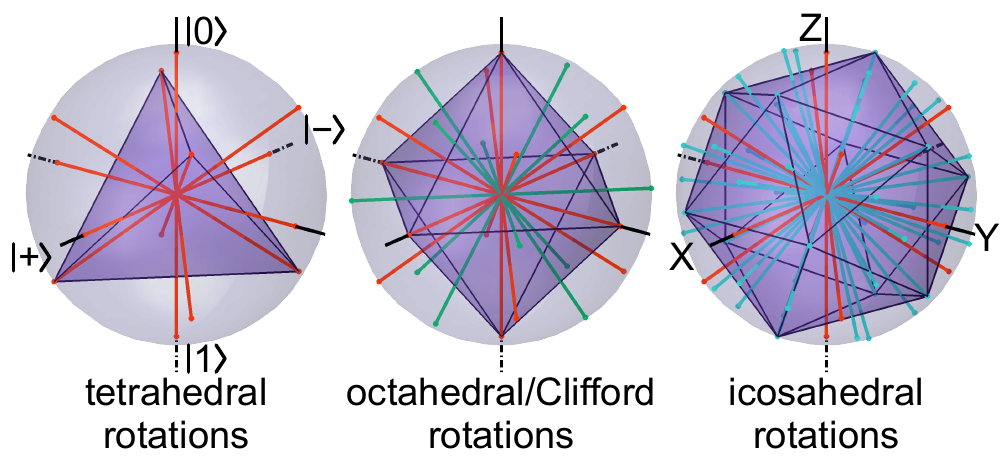}
    \caption{\textbf{The Platonic solids and their rotational groups.}
    The axes of rotation are of the tetra-, octa- and icosahedral rotational group;
    the respective Platonic solids are superimposed.
    The axes are defined by lines intersecting the origin,
    and a vertex, face center, or midpoint of an edge.
    The tetrahedral rotational group (orange) is shared among all groups.}
    \label{fig:rotgroups}
\end{figure}

\begin{figure}[t!]
    \centering
    \includegraphics[width=0.48\textwidth]{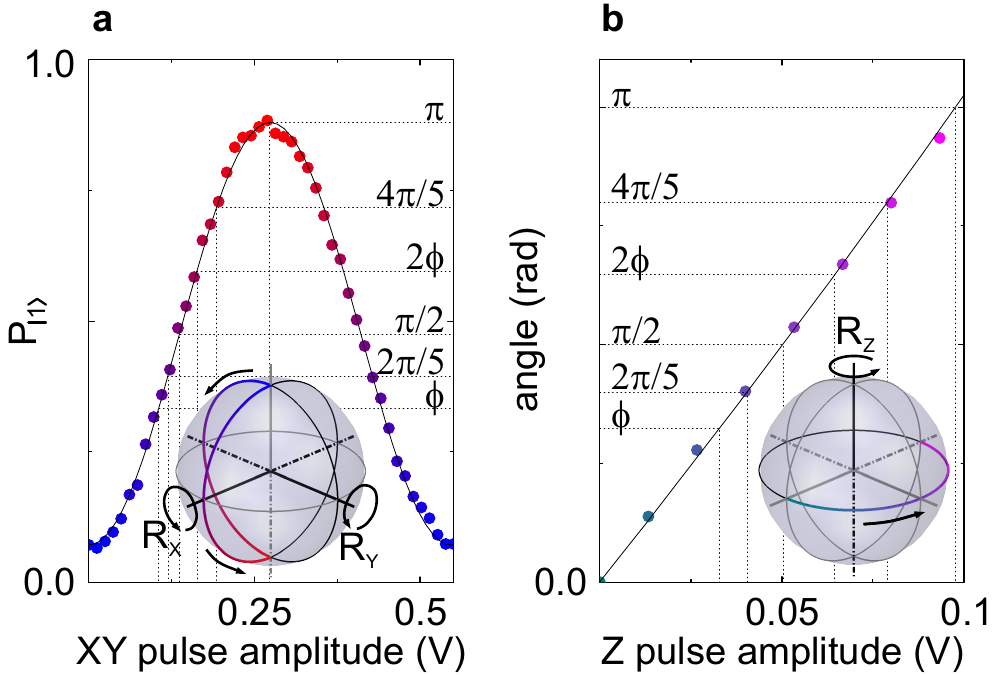}
    \caption{\textbf{Calibrating the angles of rotation.}
    (a) The excited state probability versus X and Y pulse voltage amplitude on the control board.
    The amplitudes for the required phases of rotations around the X and Y axes are indicated with dotted lines.
    The data follow a $\sin^2$ dependence (solid line) on the pulse amplitude, as expected.
    Data not corrected for measurement fidelity.
    (b) The phase of the quantum state as a function of Z pulse voltage amplitude, measured using quantum state tomography. Solid line is a fit to the data.
    For brevity, only the positive angles are shown. Here $\tan \phi=(1+\sqrt{5})/2$.
    Insets show the trajectories on the Bloch sphere for the X-, Y-, and Z-axis rotations.}
    \label{fig:tuneup}
\end{figure}

\begin{figure*}[t!]
    \centering
    \includegraphics[width=1\textwidth]{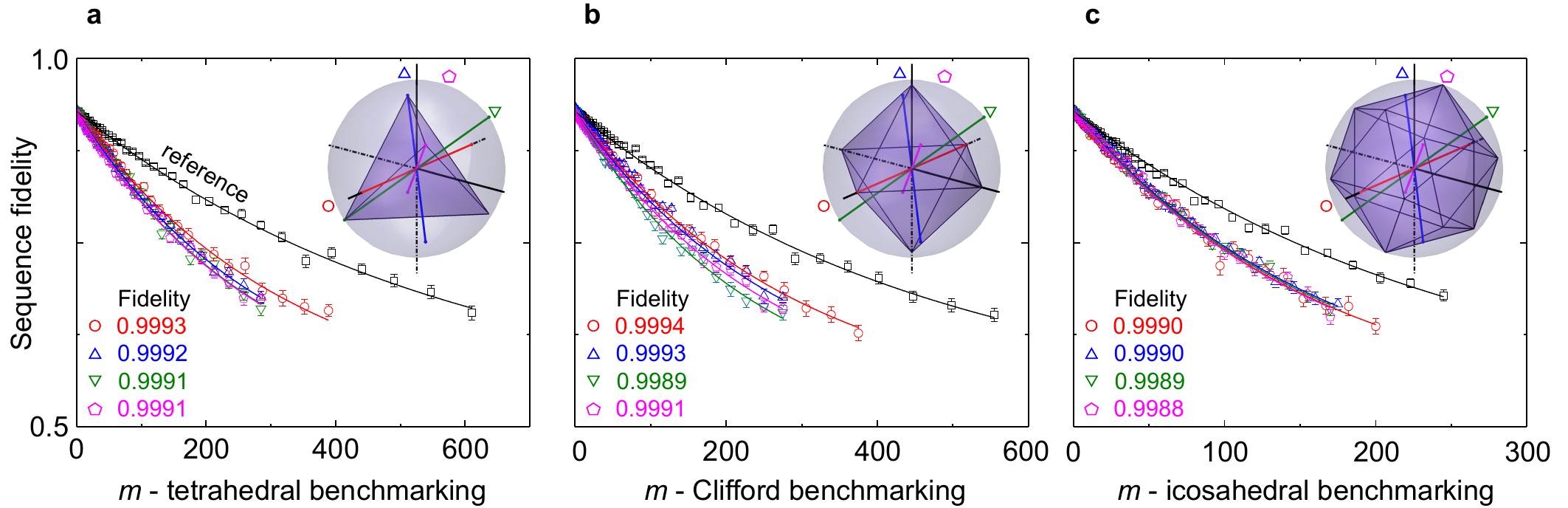}
    \caption{\textbf{Randomized benchmarking with the (a) tetra-, (b) octa- and (c) icosahedral rotational groups.}
    The sequence fidelities are plotted as a function of $m$, the
    number of random rotations or sets of random rotation and
    interleaved gate. For each $m$, the fidelity is averaged over $k=50$ different, random sequences.
    From fits to the reference curves (black lines) we extract the average error
    per group rotation of $r_\mathrm{ref,T}=0.0009$, $r_\mathrm{ref,C}=0.0010$, and
    $r_\mathrm{ref,I}=0.0019$, consistent with an average physical
    gate fidelity of $0.9995$. The rotational groups preserve Platonic solids in Bloch space,
    the respective solids are shown in the insets. The colored lower curves show the data when interleaving four tetrahedral
    rotations which are shared among all the three groups, the rotational axes are shown in the
    insets; the composed gates are X$_{\pi}$ ($\raise.08\baselineskip\hbox{\fullmoon}$), X$_{\pi/2}$ Y$_{\pi/2}$ ($\vartriangle$), X$_{-\pi/2}$
    Y$_{\pi/2}$ ($\triangledown$), and Y$_{\pi/2}$ X$_{\pi/2}$
    ($\pentagon$). Here, X$_{\pi/2}$ Y$_{\pi/2}$ denotes the unitary $R_\mathrm{Y}(\pi/2)\cdot R_\mathrm{X}(\pi/2)=\exp(-i\pi \sigma_\mathrm{Y} /4) \cdot \exp(-i\pi \sigma_\mathrm{X} /4)$.
    The gate fidelities are tabulated in the figures, extracted from fits to the data
    (solid lines). Error bars on the data indicate the standard deviation of the mean.}
    \label{fig:rball}
\end{figure*}

The groups of unitaries we use here are formed by the rotations that
preserve the regular tetrahedron, octahedron, and icosahedron --
Platonic solids -- in the Bloch sphere representation, see
Fig.~\ref{fig:rotgroups}. These are the rotational subgroups of the
tetrahedral, octahedral and icosahedral symmetry groups $T_h$, $O_h$
and $I_h$. These rotations exchange faces, amounting to a quantum
version of rolling dice (such dice are referred to as d4, d8, and
d20), but now in Bloch space. The tetrahedral, octahedral and
icosahedral rotational groups have size (order) 12, 24, and 60,
respectively. The axes are defined by the lines that intersect the
origin, and a face center, vertex, or midpoint of an edge. The angles
of rotation around these axes are, respectively, integer multiples of
$\{ 2\pi/3, 2\pi/3, \pi \}$ for the tetrahedral group, $\{ 2\pi/3,
2\pi/4, \pi \}$ for the octahedral group, and $\{ 2\pi/3, 2\pi/5, \pi
\}$ for the icosahedral group. The tetrahedral rotations (orange axes
in Fig.~\ref{fig:rotgroups}) are shared among all three groups,
enabling comparison experiments. The octahedral rotations form the
single-qubit Clifford group. The icosahedral rotations form the most
dense group -- the icosahedron is the largest of the Platonic solids
-- allowing for fine unitary control.

For implementing gates from these groups we decompose them into
rotations around the X, Y, and Z axes. The tetra- and octahedral
groups can be implemented using only $\pi/2$ and $\pi$ rotations
\cite{corcoles2013}. The icosahedral group requires the following
rotation angles: $\{ \phi, 2\pi/5, \pi/2, 2\phi, 4\pi/5,\pi \}$, with
$\phi$ an irrational angle from $\tan \phi =(1+\sqrt{5})/2$ the
golden ratio. The decomposition into physical gates is shown in the
Supplementary Information. The average number of physical gates per
tetra-, octa- or icosahedral rotation is $1\frac{3}{4}$,
$1\frac{7}{8}$, and $4\frac{4}{15}$, respectively. This decomposition
requires a minimal number of used angles and only one irrational
angle.

The rotations are implemented in our superconducting quantum system,
the Xmon transmon qubit \cite{barends2013}. This qubit combines full,
direct axial control with a high level of coherence. Details of the
device used in this experiment can be found in
Ref.~\cite{barends2014}. Rotations around the X and Y axes are
achieved by applying microwave pulses. Rotations around the Z axis
can be directly performed by detuning the qubit frequency, or by
combining X and Y rotations. All control pulses have cosine
envelopes, generated by fast (1 Gsample/sec) digital-to-analog
converter boards. For XY control we generate both the in-phase and
quadrature component and upconvert it to the qubit frequency using
quadrature mixing, see Supplementary information and
Refs.~\cite{barends2014,kelly2014} for more detail. For calibrating
the pulse amplitudes we use the measured probability for X and Y
rotations, and for Z rotations the phase as determined using quantum
state tomography (Fig.~\ref{fig:tuneup}). We minimize leakage to
energy levels above the computational subspace by applying a
quadrature correction \cite{lucero2010,chow2010}. Subsequently,
fine-tuning of the parameters is done through optimized randomized
benchmarking for immediate tune-up (ORBIT) \cite{kelly2014}, reducing
gate errors by approximately $10^{-4}$ \cite{rbimprovement}. The
generators of the tetrahedral and octahedral group are fully
parameterized by a total of three parameters, and the generators of
the icosahedral group by a total of 14 variables (Supplementary
Information).

We test the gates using randomized benchmarking
\cite{barends2014,ryan2009,harty2014,corcoles2013,magesan2011}. In
essence, randomized benchmarking is equivalent to randomly rolling
the die in Bloch space $m$ times followed by a final rotation that
returns it to the starting position, and then measuring the
probability of success. One would like to determine the gate error
averaged over all possible input states. As the gate error depends
quadratically on, for example, any amount of over- or underrotation,
we do not need to evaluate a continuum of input states. The average
of a polynomial function of order $t$ over the surface of a sphere
can be evaluated exactly using only a finite number of points, such a
group of points is a spherical $t$-design. For the single-qubit case,
unitary designs are the group of rotations that can generate
spherical designs, mapping between the points \cite{gross2007}.
Therefore, the rotational group used in randomized benchmarking needs
to be a unitary 2-design
\cite{gross2007,emerson2005,dankert2009,magesan2012pra}. Moreover,
unitary 2-designs depolarize any error in the computational basis.
For the single-qubit case, the rotational groups which preserve
Platonic solids are the 2-designs \cite{2design}. There are only
three unitary 2-designs as the cube shares the same group as the
octahedron (the cube and the octahedron are duals), and the
dodecahedron shares the same rotations as the icosahedron (the
dodecahedron and the icosahedron are duals). We have thus tested all
unitary 2-designs in Bloch space.

Randomized benchmarking with 2-designs is therefore a crucial test of
coherent control. The decrease of the probability of success -- the
sequence fidelity -- with increasing sequence length is used to
quantify the gate fidelity. We start by measuring a reference curve,
using sequences of $m$ random rotations. The sequence fidelity
follows $A p^m + B$, with variables $A$ and $B$ absorbing measurement
and initialization errors, and $p_\mathrm{ref}$ giving the average
error per rotation: $r_\mathrm{ref}=(1-p_\mathrm{ref})/2$
\cite{magesan2011}. We then interleave a specific gate with $m$
random rotations, the difference with the reference is a direct
measure of the gate error:
$r_\mathrm{gate}=(1-p_\mathrm{gate}/p_\mathrm{ref})/2$, the gate
fidelity is $F_\mathrm{gate}=1-r_\mathrm{gate}$ \cite{magesan2012}.
At each $m$, the data is averaged over $k=50$ random sequences
\cite{epstein2013}.

We have performed randomized benchmarking using the tetrahedral,
octahedral, and icosahedral rotational groups; the results are shown
in Fig.~\ref{fig:rball}. As we start by initializing the qubit in the
ground state, the sequence fidelity is given by the ground state
population after applying the random sequences. The traces follow an
exponential decay with increasing $m$, as expected. We have also
interleaved four gates from the tetrahedral group (see insets for the
rotational axes). These rotations are shared by all three rotational
groups, allowing for a direct comparison between tetra-, octa-, and
icosahedral-based randomized benchmarking. We emphasize that the
interleaved gates are physically implemented in exactly the same
manner.

From the reference traces, we extract an average error per group of
rotations of $r_\mathrm{ref,T}=9\cdot 10^{-4}$,
$r_\mathrm{ref,C}=10\cdot 10^{-4}$, $r_\mathrm{ref,I}=19\cdot
10^{-4}$. When dividing by $1\frac{3}{4}$, $1\frac{7}{8}$ or
$4\frac{4}{15}$, these numbers consistently point to an average error
of $5\cdot 10^{-4}$ per physical gate (single decomposed rotation
around the X, Y, or Z axis). The extracted fidelities for the
interleaved gates are tabulated in Fig.~\ref{fig:rball}. The
reference error per gate, as well as the errors for the interleaved
gates, are consistent with previous measurements \cite{barends2014},
where the average physical gate fidelity lies at 0.9994. In addition,
the mean difference in error of the interleaved gates is below $2
\cdot 10^{-4}$, verifying that any of the groups can be used for
randomized benchmarking.

\begin{figure}[b!]
    \centering
    \includegraphics[width=0.48\textwidth]{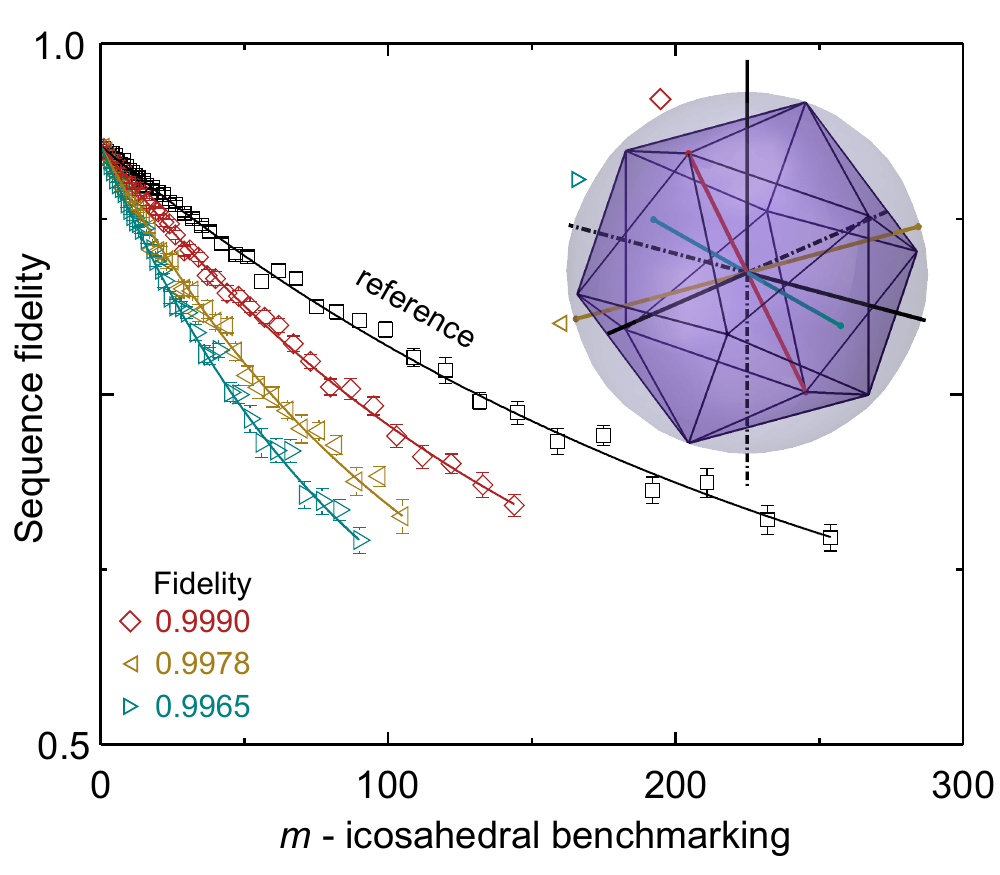}
    \caption{\textbf{Icosahedral-based randomized benchmarking.}
    We have interleaved three non-Clifford gates whose axes are shown in the inset, the gates rotate around a face center, vertex or edge midpoint of
    the icosahedron (superimposed). The gates are composed of three, six, and eight elements. Their compositions are: Y$_{\phi}$ X$_{2\pi/5}$ Y$_{-\phi}$ ($\diamond$),
    X$_{\phi}$ Z$_{-2\pi/5}$ Y X$_{2\phi}$ Z$_{2\pi/5}$ X$_{-\phi}$ ($\vartriangleleft$),
    and X$_{\phi}$ Z$_{-2\pi/5}$ X$_{-\phi}$ X$_{-\pi/2}$ Y$_{-\pi/2}$ X$_{\phi}$ Z$_{2\pi/5}$ X$_{-\phi}$ ($\vartriangleright$).
    The gate fidelities are tabulated in the figure.
    The average error per physical gate which makes up the interleaved gates is $r=3-4 \cdot 10^{-4}$.}
    \label{fig:rbico}
\end{figure}

With icosahedral randomized benchmarking shown to be a viable method
for determining gate fidelity, we can now benchmark gates outside of
the Clifford group, as shown in Fig.~\ref{fig:rbico}. We chose three
composite gates, which are implemented using three, six, or eight
physical gates. The rotational axes are highlighted in the inset. The
fidelities of these gates are tabulated in the figure. These complex
gates work surprisingly well: we compute the average error per
physical decomposition to range between $3 \cdot 10^{-4}$ and $4\cdot
10^{-4}$, assuming that errors are small and uncorrelated. These
results demonstrate that even these complex, composite gates, can be
implemented with high fidelity.

Apart from the first demonstrated implementation of rotational groups
beyond the Clifford group, the results on icosahedral benchmarking in
Figs.~\ref{fig:rball} and \ref{fig:rbico} clearly indicate that
physical rotations, other than the widely used Clifford rotations,
can be done with a very similar fidelity. This strongly suggests that
any arbitrary rotation can be done with high fidelity. Moreover, the
gate parameters can be optimized to achieve decoherence-limited
performance using the method outlined in Ref.~\cite{kelly2014},
providing an interpolation table for implementing any desired
rotation directly, efficiently, and accurately. In addition,
icosahedral benchmarking could also be used for evaluating functions
of higher order, beyond gate fidelity, as the tetra-, octa-, and
icosahedral rotational groups are unitary 2-, 3-, and 5-designs
\cite{roy2009,pauli}.

We have shown a quantum version of rolling dice with a
superconducting qubit, using gate sets inspired by the Platonic
solids. Fundamentally, our work illustrates the potential of using
unitaries with a finer distribution for accurate control, and
provides a route for the implementation and benchmarking of
non-Clifford gates. More generally, our results imply that arbitrary
rotations can be done with high accuracy, allowing for complex gates
and algorithms to be performed more efficiently in quantum
information processing.

\smallskip\noindent
\textbf{Acknowledgements} R.B. acknowledges G. P. Velders for
exhausting demonstrations of rolling dice. This research was funded
by the Office of the Director of National Intelligence (ODNI),
Intelligence Advanced Research Projects Activity (IARPA), through the
Army Research Office grants W911NF-09-1-0375 and W911NF-10-1-0334.
All statements of fact, opinion or conclusions contained herein are
those of the authors and should not be construed as representing the
official views or policies of IARPA, the ODNI, or the U.S.
Government. Devices were made at the UC Santa Barbara Nanofabrication
Facility, a part of the NSF-funded National Nanotechnology
Infrastructure Network, and at the NanoStructures Cleanroom Facility.

\smallskip\noindent
\textbf{Author contributions} 
R.B. and J.K designed the sample and performed the experiment. R.B.
and A.V. designed the experiment. J.K., A.M., and R.B. fabricated the
sample. R.B., J.K., A.V., A.G.F., A.N.K., and J.M.M. co-wrote the
manuscript. All authors contributed to the fabrication process,
experimental set-up, and manuscript preparation.

\smallskip\noindent
\textbf{Additional information} The authors declare no competing
financial interests. Supplementary information accompanies this paper
on [weblink to be inserted by editor]. Reprints and permissions
information is available at [weblink to be inserted by editor].
Correspondence and requests for materials should be addressed to R.B.
or J.M.M.


\begin{thebibliography}{99}
\bibitem{rbimprovement} We find that fine-tuning with randomized
    benchmaring reduces the average error from
    $10\cdot 10^{-4}$ to $9\cdot 10^{-4}$ per Clifford and $20\cdot
    10^{-4}$ to $19\cdot 10^{-4}$ per icosahedral rotation.
\end{thebibliography}
\end{document}


\title{Supplementary Information for ``Rolling quantum dice with a superconducting qubit''}

\author{R. Barends}
\thanks{These authors contributed equally to this work.}
\author{J. Kelly}
\thanks{These authors contributed equally to this work.}
\affiliation{Department of Physics, University of California, Santa
Barbara, CA 93106, USA}
\author{A. Veitia}
\affiliation{Department of Electrical Engineering, University of
California, Riverside, CA 92521, USA}
\author{A. Megrant}
\affiliation{Department of Physics, University of California, Santa
Barbara, CA 93106, USA}
\author{A. G. Fowler}
\affiliation{Department of Physics, University of California, Santa
Barbara, CA 93106, USA} \affiliation{Centre for Quantum Computation
and Communication Technology, School of Physics, The University of
Melbourne, Victoria 3010, Australia}

\author{B. Campbell}
\author{Y. Chen}
\author{Z. Chen}
\author{B. Chiaro}
\author{A. Dunsworth}
\author{I.-C. Hoi}
\author{E. Jeffrey}
\author{C. Neill}
\author{P. J. J. O'Malley}
\author{J. Mutus}
\author{C. Quintana}
\author{P. Roushan}
\author{D. Sank}
\author{J. Wenner}
\author{T. C. White}
\affiliation{Department of Physics, University of California, Santa
Barbara, CA 93106, USA}

\author{A. N. Korotkov}
\affiliation{Department of Electrical Engineering, University of
California, Riverside, CA 92521, USA}
\author{A. N. Cleland}
\author{John M. Martinis}
\affiliation{Department of Physics, University of California, Santa
Barbara, CA 93106, USA}

\maketitle

\section{Rotational groups}

The tetra-, octa-, and icosahedral rotational groups are shown in
Tables.~\ref{table:tetra}, \ref{table:octa} and \ref{table:ico}.

\begin{table*}[b!]
\centering \caption{The tetrahedral rotational group, written in
terms of the physical microwave gates applied in time. Negative
angles are included through opposite rotational axes.}
\begin{tabular}{  c| c } 
    \hline
    \hline

    \begin{tabular}[t]{ll}
        \multicolumn{2}{c}{Paulis - $\pi$}\\
        \hline
        I& \\
        X$_{\pi}$& \\
        Y$_{\pi}$& \\
        Y$_{\pi}$&X$_{\pi}$ \\
    \end{tabular} &

    \begin{tabular}[t]{ll}
        \multicolumn{2}{c}{$2\pi/3$}\\
        \hline
        X$_{\pi/2}$&Y$_{\pi/2}$ \\
        X$_{\pi/2}$&Y$_{-\pi/2}$ \\
        X$_{-\pi/2}$&Y$_{\pi/2}$ \\
        X$_{-\pi/2}$&Y$_{-\pi/2}$ \\
        Y$_{\pi/2}$&X$_{\pi/2}$ \\
        Y$_{\pi/2}$&X$_{-\pi/2}$ \\
        Y$_{-\pi/2}$&X$_{\pi/2}$ \\
        Y$_{-\pi/2}$&X$_{-\pi/2}$ \\
    \end{tabular} \\

    \hline \hline
\end{tabular}
\label{table:tetra}
\end{table*}

\begin{table*}[b!]
\centering \caption{The octahedral rotational group -- single qubit
Cliffords. The Paulis and $2\pi/3$ rotations form the tetrahedral
rotational group.}
\begin{tabular}{  c| c| c| c } 
    \hline
    \hline
    \begin{tabular}[t]{ll}
        \multicolumn{2}{c}{Paulis - $\pi$}\\
        \hline
        I& \\
        X$_{\pi}$& \\
        Y$_{\pi}$& \\
        Y$_{\pi}$&X$_{\pi}$ \\
    \end{tabular} &

    \begin{tabular}[t]{ll}
        \multicolumn{2}{c}{$2\pi/3$}\\
        \hline
        X$_{\pi/2}$&Y$_{\pi/2}$ \\
        X$_{\pi/2}$&Y$_{-\pi/2}$ \\
        X$_{-\pi/2}$&Y$_{\pi/2}$ \\
        X$_{-\pi/2}$&Y$_{-\pi/2}$ \\
        Y$_{\pi/2}$&X$_{\pi/2}$ \\
        Y$_{\pi/2}$&X$_{-\pi/2}$ \\
        Y$_{-\pi/2}$&X$_{\pi/2}$ \\
        Y$_{-\pi/2}$&X$_{-\pi/2}$ \\
    \end{tabular} &

    \begin{tabular}[t]{lll}
        \multicolumn{3}{c}{$\pi/2$}\\
        \hline
        X$_{\pi/2}$&& \\
        X$_{-\pi/2}$&& \\
        Y$_{\pi/2}$&& \\
        Y$_{-\pi/2}$&& \\
        X$_{-\pi/2}$&Y$_{\pi/2}$&X$_{\pi/2}$ \\
        X$_{-\pi/2}$&Y$_{-\pi/2}$&X$_{\pi/2}$ \\
    \end{tabular} &

    \begin{tabular}[t]{lll}
        \multicolumn{3}{c}{Hadamard-like - $\pi$}\\
        \hline
        X$_{\pi}$&Y$_{\pi/2}$& \\
        X$_{\pi}$&Y$_{-\pi/2}$& \\
        Y$_{\pi}$&X$_{\pi/2}$& \\
        Y$_{\pi}$&X$_{-\pi/2}$& \\
        X$_{\pi/2}$&Y$_{\pi/2}$&X$_{\pi/2}$ \\
        X$_{-\pi/2}$&Y$_{\pi/2}$&X$_{-\pi/2}$ \\
    \end{tabular} \\

    \hline \hline
\end{tabular}
\label{table:octa}
\end{table*}

\begin{table*}[b!]
\centering \caption{The 60 icosahedral rotations, excluding the idle
(I). The rotations are ordered based on their angles, and their
points of intersection with the icosahedron. The edges and faces
contain the Paulis and $2\pi/3$ rotations which overlap with the
tetrahedral rotational group. For the edge rotations we have used
$R_\mathrm{X}(\phi) R_\mathrm{Y}(\pi) R_\mathrm{X}(-\phi) =
R_\mathrm{X}(2\phi) R_\mathrm{Y}(\pi)$ and $R_\mathrm{X}(\phi)
R_\mathrm{Z}(\pi) R_\mathrm{X}(-\phi) = R_\mathrm{X}(2\phi)
R_\mathrm{Z}(\pi)$ to reduce the gate count.}
\begin{tabular}{  c| c| c} 
    \hline
    \hline
    \begin{tabular}[t]{lll}
        \multicolumn{3}{c}{Vertices - $2\pi/5$}\\
        \hline
        Y$_{\phi}$&X$_{2\pi/5}$&Y$_{-\phi}$ \\        
        Y$_{\phi}$&X$_{-2\pi/5}$&Y$_{-\phi}$ \\
        Y$_{-\phi}$&X$_{2\pi/5}$&Y$_{\phi}$ \\
        Y$_{-\phi}$&X$_{-2\pi/5}$&Y$_{\phi}$ \\
        Z$_{\phi}$&Y$_{2\pi/5}$&Z$_{-\phi}$ \\
        Z$_{\phi}$&Y$_{-2\pi/5}$&Z$_{-\phi}$ \\
        Z$_{-\phi}$&Y$_{2\pi/5}$&Z$_{\phi}$ \\
        Z$_{-\phi}$&Y$_{-2\pi/5}$&Z$_{\phi}$ \\
        X$_{\phi}$&Z$_{2\pi/5}$&X$_{-\phi}$ \\
        X$_{\phi}$&Z$_{-2\pi/5}$&X$_{-\phi}$ \\
        X$_{-\phi}$&Z$_{2\pi/5}$&X$_{\phi}$ \\
        X$_{-\phi}$&Z$_{-2\pi/5}$&X$_{\phi}$ \\
        \hline
        \hline
        \multicolumn{3}{c}{Vertices - $4\pi/5$}\\
        \hline
        Y$_{\phi}$&X$_{4\pi/5}$&Y$_{-\phi}$ \\
        Y$_{\phi}$&X$_{-4\pi/5}$&Y$_{-\phi}$ \\
        Y$_{-\phi}$&X$_{4\pi/5}$&Y$_{\phi}$ \\
        Y$_{-\phi}$&X$_{-4\pi/5}$&Y$_{\phi}$ \\
        Z$_{\phi}$&Y$_{4\pi/5}$&Z$_{-\phi}$ \\
        Z$_{\phi}$&Y$_{-4\pi/5}$&Z$_{-\phi}$ \\
        Z$_{-\phi}$&Y$_{4\pi/5}$&Z$_{\phi}$ \\
        Z$_{-\phi}$&Y$_{-4\pi/5}$&Z$_{\phi}$ \\
        X$_{\phi}$&Z$_{4\pi/5}$&X$_{-\phi}$ \\
        X$_{\phi}$&Z$_{-4\pi/5}$&X$_{-\phi}$ \\
        X$_{-\phi}$&Z$_{4\pi/5}$&X$_{\phi}$ \\
        X$_{-\phi}$&Z$_{-4\pi/5}$&X$_{\phi}$ \\
    \end{tabular} &

    \begin{tabular}[t]{llllllll}
        \multicolumn{8}{c}{Faces - $2\pi/3$}\\
        \hline
        X$_{-\pi/2}$&Y$_{-\pi/2}$&&&&&& \\
        Y$_{\pi/2}$&X$_{\pi/2}$&&&&&& \\
        X$_{\phi}$&Z$_{-2\pi/5}$&X$_{-\phi}$&X$_{-\pi/2}$&Y$_{-\pi/2}$&X$_{\phi}$&Z$_{2\pi/5}$&X$_{-\phi}$ \\    
        X$_{\phi}$&Z$_{-2\pi/5}$&X$_{-\phi}$&Y$_{\pi/2}$&X$_{\pi/2}$&X$_{\phi}$&Z$_{2\pi/5}$&X$_{-\phi}$ \\
        X$_{\phi}$&Z$_{-4\pi/5}$&X$_{-\phi}$&X$_{-\pi/2}$&Y$_{-\pi/2}$&X$_{\phi}$&Z$_{4\pi/5}$&X$_{-\phi}$ \\
        X$_{\phi}$&Z$_{-4\pi/5}$&X$_{-\phi}$&Y$_{\pi/2}$&X$_{\pi/2}$&X$_{\phi}$&Z$_{4\pi/5}$&X$_{-\phi}$ \\
        X$_{-\pi/2}$&Y$_{\pi/2}$&&&&&& \\
        Y$_{-\pi/2}$&X$_{\pi/2}$&&&&&& \\
        X$_{\phi}$&Z$_{2\pi/5}$&X$_{-\phi}$&X$_{-\pi/2}$&Y$_{-\pi/2}$&X$_{\phi}$&Z$_{-2\pi/5}$&X$_{-\phi}$ \\
        X$_{\phi}$&Z$_{2\pi/5}$&X$_{-\phi}$&Y$_{\pi/2}$&X$_{\pi/2}$&X$_{\phi}$&Z$_{-2\pi/5}$&X$_{-\phi}$ \\
        X$_{\pi/2}$&Y$_{\pi/2}$&&&&&& \\
        Y$_{-\pi/2}$&X$_{-\pi/2}$&&&&&& \\
        X$_{\phi}$&Z$_{-4\pi/5}$&X$_{-\phi}$&X$_{\pi/2}$&Y$_{\pi/2}$&X$_{\phi}$&Z$_{4\pi/5}$&X$_{-\phi}$ \\
        X$_{\phi}$&Z$_{-4\pi/5}$&X$_{-\phi}$&Y$_{-\pi/2}$&X$_{-\pi/2}$&X$_{\phi}$&Z$_{4\pi/5}$&X$_{-\phi}$ \\
        X$_{\phi}$&Z$_{4\pi/5}$&X$_{-\phi}$&X$_{\pi/2}$&Y$_{\pi/2}$&X$_{\phi}$&Z$_{-4\pi/5}$&X$_{-\phi}$ \\
        X$_{\phi}$&Z$_{4\pi/5}$&X$_{-\phi}$&Y$_{-\pi/2}$&X$_{-\pi/2}$&X$_{\phi}$&Z$_{-4\pi/5}$&X$_{-\phi}$ \\
        X$_{\phi}$&Z$_{2\pi/5}$&X$_{-\phi}$&X$_{\pi/2}$&Y$_{\pi/2}$&X$_{\phi}$&Z$_{-2\pi/5}$&X$_{-\phi}$ \\
        X$_{\phi}$&Z$_{2\pi/5}$&X$_{-\phi}$&Y$_{-\pi/2}$&X$_{-\pi/2}$&X$_{\phi}$&Z$_{-2\pi/5}$&X$_{-\phi}$ \\
        X$_{\pi/2}$&Y$_{-\pi/2}$&&&&&& \\
        Y$_{\pi/2}$&X$_{-\pi/2}$&&&&&& \\
    \end{tabular} &

    \begin{tabular}[t]{llllll}
        \multicolumn{6}{c}{Edges - $\pi$}\\
        \hline
        X$_{\pi}$&&&&& \\
        X$_{\phi}$&Z$_{2\pi/5}$&X$_{\pi}$&Z$_{-2\pi/5}$&X$_{-\phi}$& \\
        X$_{\phi}$&Z$_{-2\pi/5}$&X$_{\pi}$&Z$_{2\pi/5}$&X$_{-\phi}$& \\
        X$_{\phi}$&Z$_{4\pi/5}$&X$_{\pi}$&Z$_{-4\pi/5}$&X$_{-\phi}$& \\
        X$_{\phi}$&Z$_{-4\pi/5}$&X$_{\pi}$&Z$_{4\pi/5}$&X$_{-\phi}$& \\
        Y$_{\pi}$&&&&& \\
        X$_{\phi}$&Z$_{2\pi/5}$&Y$_{\pi}$&X$_{2\phi}$&Z$_{-2\pi/5}$&X$_{-\phi}$ \\
        X$_{\phi}$&Z$_{-2\pi/5}$&Y$_{\pi}$&X$_{2\phi}$&Z$_{2\pi/5}$&X$_{-\phi}$ \\      
        X$_{\phi}$&Z$_{4\pi/5}$&Y$_{\pi}$&X$_{2\phi}$&Z$_{-4\pi/5}$&X$_{-\phi}$ \\
        X$_{\phi}$&Z$_{-4\pi/5}$&Y$_{\pi}$&X$_{2\phi}$&Z$_{4\pi/5}$&X$_{-\phi}$ \\
        Z$_{\pi}$&&&&& \\
        X$_{\phi}$&Z$_{2\pi/5}$&Z$_{\pi}$&X$_{2\phi}$&Z$_{-2\pi/5}$&X$_{-\phi}$ \\
        X$_{\phi}$&Z$_{-2\pi/5}$&Z$_{\pi}$&X$_{2\phi}$&Z$_{2\pi/5}$&X$_{-\phi}$ \\
        X$_{\phi}$&Z$_{4\pi/5}$&Z$_{\pi}$&X$_{2\phi}$&Z$_{-4\pi/5}$&X$_{-\phi}$ \\
        X$_{\phi}$&Z$_{-4\pi/5}$&Z$_{\pi}$&X$_{2\phi}$&Z$_{4\pi/5}$&X$_{-\phi}$ \\
    \end{tabular} \\

    \hline \hline
\end{tabular}
\label{table:ico}
\end{table*}

\section{Pulse calibration}

The microwave pulses -- the rotations around the X and Y axes -- are
created by generating envelopes using 1 Gsample/s digital to analog
converters, and upconverting these envelopes to the qubit frequency
using quadrature mixing (see Ref.~\cite{barends2014} for the control
and readout system). The room temperature electronics are calibrated
by corrected the pulse from distortion using deconvolution
techniques, and by correcting the quadrature mixer for gain and phase
imbalances. The pulses for frequency control -- the rotations around
the Z axis -- are generated by 1 Gsample/s digital to analog
converters; non-idealities in the pulse shape from room temperature
electronics are suppressed by deconvolution techniques.
Non-idealities arising from stray inductance and reflections in the
wiring of the cryostat are suppressed by using the qubit to measure
the step response and by randomized benchmarking, see
Refs.~\cite{barends2014,kelly2014} for details.

We then use the qubit to calibrate the pulse amplitudes, and the DRAG
(derivative reduction for adiabatic gates) parameter for minimizing
2-state leakage \cite{lucero2010,chow2010}. We do not calibrate the
phase between a X and Y rotation using the qubit
\cite{gustavsson2012} as the quadrature mixer calibrations are
sufficient. The pulse amplitudes for Z rotations are determined using
quantum state tomography.

We use three parameters to generate the microwave pulses necessary
for the tetrahedral and octahedral rotational groups (DRAG parameter,
two pulse amplitudes) and 14 parameters (DRAG parameter, 13 pulse
amplitudes) to generate the pulses for the icosahedral rotational
group; see Table~\ref{table:generators} for the generators.

\begin{table*}[b!]
\centering \caption{Generators of the tetra-, octa-, and icosahedral
rotational groups, excluding the idle. The generators are listed in
terms of shared pulse amplitude parameter. Including the DRAG
parameter, we use a total of three parameters to generate the tetra-,
and octahedral rotational group (DRAG parameter, pulse amplitude
parameter for  X$_{\pi}$ and Y$_{\pi}$, pulse amplitude parameter for
X$_{\pi/2}$, X$_{-\pi/2}$, Y$_{\pi/2}$, and Y$_{-\pi/2}$), and a
total of 14 parameters to generate the icosahedral rotational group.
The duration of the idle and each of the X and Y gates is 12~ns, and
each Z gate is 10~ns.}
\begin{tabular}{  c | c } 
    \hline
    \hline
    Rotational group & Generators \\
    \hline

    \multirow{2}{*}{Tetrahedral, octahedral, and icosahedral}

    &
    \begin{tabular}[t]{llll}
        X$_{\pi}$, & Y$_{\pi}$ & & \\
        X$_{\pi/2}$, & X$_{-\pi/2}$, & Y$_{\pi/2}$, & Y$_{-\pi/2}$
    \end{tabular}
    \\
    \hline

    \multirow{11}{*}{Icosahedral}
    &
    \begin{tabular}[t]{llll}
        X$_{2\pi/5}$, & X$_{-2\pi/5}$, & Y$_{2\pi/5}$, & Y$_{-2\pi/5}$ \\
        X$_{4\pi/5}$, & X$_{-4\pi/5}$, & Y$_{4\pi/5}$, & Y$_{-4\pi/5}$ \\
        X$_{\phi}$, & Y$_{\phi}$, & X$_{-\phi}$, & Y$_{-\phi}$\\
        X$_{2\phi}$ & & &\\
        Z$_{2\pi/5}$ & & &\\
        Z$_{-2\pi/5}$ & & &\\
        Z$_{\phi}$ & & &\\
        Z$_{-\phi}$  & & &\\
        Z$_{4\pi/5}$ & & &\\
        Z$_{-4\pi/5}$ & & &\\
        Z$_{\pi}$ & & &
    \end{tabular}\\

    \hline \hline

\end{tabular}
\label{table:generators}
\end{table*}